# Visible, EUV, and X-ray Spectroscopy at the NIST EBIT Facility

J. D. Gillaspy[1], B. Blagojevic[1], A. Dalgarno[2], K. Fahey[1,3], V. Kharchenko[2], J. M. Laming[4], E.-O. Le Bigot[1], L. Lugosi[5], K. Makonyi[1], L. P. Ratliff[1], H. W. Schnopper[2], E. H. Silver[2], E. Takács[1,6,7], J. N. Tan[1], H. Tawara[1,8], K. Tokési[5]

[1]*National Institute of Standards and Technology, Gaithersburg, MD 20899-8421, USA*
[2]*Harvard-Smithonian Center for Astrophysics, 60 Garden St., Cambridge, MA 0213, USA*
[3]*University College Dublin, Department of Experimental Physics. Belfield, Dublin 4, Ireland*
[4]*E. O. Hurlburt Center for Space Research, Naval Research Lab, Washington, DC 20375, USA*
[5]*Institute of Nuclear Research of the Hungarian Academy of Sciences, Debrecen, Bem tér 18/c, Hungary, H-4026*
[6]*University of Debrecen, Experimental Physics Department, Debrecen, Bem tér 18/a, Hungary, H-4026*
[7]*Massachusetts Institute of Technology, 77 Massachusetts Avenue, Cambridge, MA 20375, USA*
[8]*The Queen's University, Department of Pure and Applied Physics, Belfast BT7 1NN, UK*

# Visible, EUV, and X-ray Spectroscopy at the NIST EBIT Facility

J. D. Gillaspy[1], B. Blagojevic[1], A. Dalgarno[2], K. Fahey[1,3], V. Kharchenko[2], J. M. Laming[4], E.-O. Le Bigot[1], L. Lugosi[5], K. Makonyi[1], L. P. Ratliff[1], H. W. Schnopper[2], E. H. Silver[2], E. Takács[1,6,7], J. N. Tan[1], H. Tawara[1,8], K. Tökési[5]

[1]*National Institute of Standards and Technology, Gaithersburg, MD 20899-8421, USA*
[2]*Harvard-Smithonian Center for Astrophysics, 60 Garden St., Cambridge, MA 0213, USA*
[3]*University College Dublin, Department of Experimental Physics. Belfield, Dublin 4, Ireland*
[4]*E. O. Hurlburt Center for Space Research, Naval Research Lab, Washington, DC 20375, USA*
[5]*Institute of Nuclear Research of the Hungarian Academy of Sciences, Debrecen, Bem tér 18/c, Hungary, H-4026*
[6]*University of Debrecen, Experimental Physics Department, Debrecen, Bem tér 18/a, Hungary, H-4026*
[7]*Massachusetts Institute of Technology, 77 Massachusetts Avenue, Cambridge, MA 20375, USA*
[8]*The Queen's University, Department of Pure and Applied Physics, Belfast BT7 1NN, UK*

**Abstract.** After a brief introduction to the NIST EBIT facility, we present the results of three different types of experiments that have been carried out there recently: EUV and visible spectroscopy in support of the microelectronics industry, laboratory astrophysics using an x-ray microcalorimeter, and charge exchange studies using extracted beams of highly charged ions.

## THE NIST EBIT FACILITY: HISTORY AND OVERVIEW

Electron Beam Ion Traps (EBITs) emerged on the world scene in 1988 with the publication of the first results by the Livermore-LBL collaboration [1]. Results that depended critically on high-resolution x-ray spectroscopy with an EBIT appeared in the Physical Review in 1989 [2], using an NRL-NIST spectrometer at Livermore, in collaboration with P. Beiersdorfer. This work followed some other important EBIT high-resolution spectroscopy work presented earlier in various conference proceedings. Two years later, construction of a new EBIT laboratory at NIST began, in collaboration with NRL, and two years after that (1993) we began trapping ions and publishing papers from the NIST facility. A compilation of the early Livermore EBIT papers was published in 1992 [3]. Recently, we published a bound collection of the NIST EBIT publications through 2001, which is available in hardcopy and CD from NIST [4]. The NIST compilation also contains some previously unpublished historical information about both the NIST and LLNL EBITs (written by the first author of the present paper, and by Ross Marrs of LLNL).

The NIST EBIT incorporates a number of design changes that are intended to allow it to reach higher beam energies than those obtained in the EBIT-I and EBIT-II at

LLNL. The upper energy limit of the NIST machine has not yet been tested, although it has been run as high as 33 keV as well as below 100 eV. We have produced charge states as high q=73+ in bismuth ions, and at 33 keV should be able to produce strong spectra from any Ne-like ion on the periodic table. The drift tube structure is not expected to hold more than 40 keV, but the upper range of the beam energy might exceed 50 keV if the electron gun and cathode are floated to negative high voltages, as they are in the LLNL SuperEBIT [5]. The NIST EBIT electron gun and collector rest on small internal high voltage platforms to enable this mode of operation, although there has been so much physics to explore below 30 keV to date that the platforms have only been run at ground potentials. For reference, we note that 33 keV is equal to the threshold energy needed to produce He-like U90+ ions, although we would not expect to produce many such ions near threshold, and have not tried to do so. A SuperEBIT of the sort available at LLNL [5], Tokyo [6], or Heidelberg [7], with energies near 200 keV, is necessary in order to produce significant quantities of the highest few percent of all the possible charge states on the periodic table.

The NIST EBIT facility is also equipped with an ion-extraction beamline for ion-solid and ion-gas collision studies. The general technique of extracting ions from an EBIT was pioneered at Livermore [8], but the design of the NIST beamline resulted in on-target beam currents of $Xe^{44+}$ ions that were 100 times higher than previously reported from an EBIT (1000x times higher in continuous beam mode) [9]. This substantial improvement (reaching 10's of pA) enabled new classes of experiments with highly charged ion beams (see, e.g., the section on charge exchange below).

## EUV AND VISIBLE SPECTROSCOPY OF XENON

In support of the development of EUV Lithography (EUVL) [10] light sources for the microelectronics industry, we have produced benchmark EUV spectra of Xe ions for testing plasma models. Xe charge states around q=10+ emit strongly around 13.5 nm wavelength, where highly efficient multilayer optics have been developed for lithography applications. The lines in this region of the spectrum appear densely packed, even when observed with the world's largest (10.7 m) grazing incidence spectrometer, located at NIST [11]. To study the intensity envelope of this group of lines (as well as the even stronger group around 11 nm) with the NIST EBIT, we have constructed an ultrahigh vacuum EUV spectrometer, based on a variable line space (VLS) flat-field grating. Figure 1 shows an example of one of our early EUV spectra.

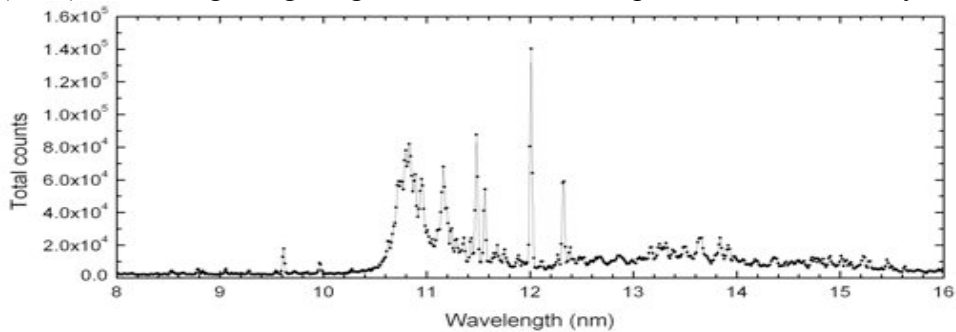

**FIGURE 1.** EUV spectrum of Xe taken at 1.5 keV electron beam energy.

To complement these EUV studies, we have also observed Xe in the x-ray and, as shown in figure 2, the visible region of the spectrum. The two lines marked on the left side of this spectrum demonstrate the improvement in our spectroscopy system since our 1995 publication that reported the first visible/uv spectroscopy on an EBIT [12]. In comparison to figure 1 of our earlier work [12], we can now obtain spectra with much higher resolution, much higher signal-to-noise ratio, and much broader spectral range, in a fraction of the time.

We note that the strongest line in figure 2 (near 598 nm) is particularly interesting in that it arises from the charge state q=9+ that is adjacent to the q=10+ charge state that future EUVL systems might be based on in the microelectronics industry. We propose that this line might be useful for real-time diagnostics of EUVL light sources.

For completeness, we note that the Livermore EBIT group has previously reported a line that is essentially at the same wavelength as the one we measure at 598 nm, but which they identify as arising from a much higher charge state (q=31+) [13]. We have tried to search for a possible coincidental line overlap by scanning the electron beam energy and looking for a slight shift in measured wavelength or intensity, but we have found no evidence for the existence of two lines at this location. The line position reported by the LLNL group is at 598.40(100) nm and is identified to be from a $3d^5$ $^4G_{9/2}$- $^4G_{11/2}$ transition in $Xe^{31+}$, while our observed line position is at 597.98(50) nm and arises from the $4d^9$ $^2D_{3/2}$- $^2D_{5/2}$ transition in $Xe^{9+}$. We can be certain that the line we see does not arise from q=31+ because it remains strong at 500 eV beam energy (fig 3a), nearly a factor of 4 below the 1.8 keV energy needed to produce $Xe^{31+}$. Further evidence supporting our identification of the charge state is provided by the fact that the line disappears when the beam energy is dropped below the 180 eV energy needed to produce $Xe^{9+}$ (fig. 3b). These measurements illustrate the power of an EBIT, when put to full use, to aid in the definitive identification of spectral lines.

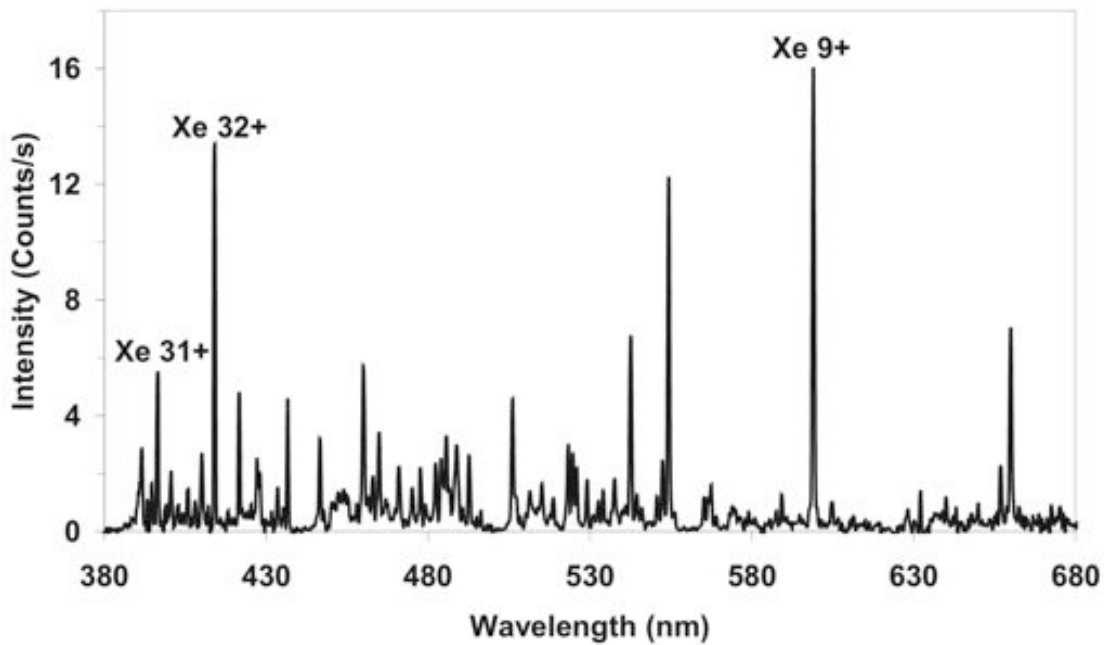

**FIGURE 2.** Spectrum of Xe taken at 3 keV electron beam energy.

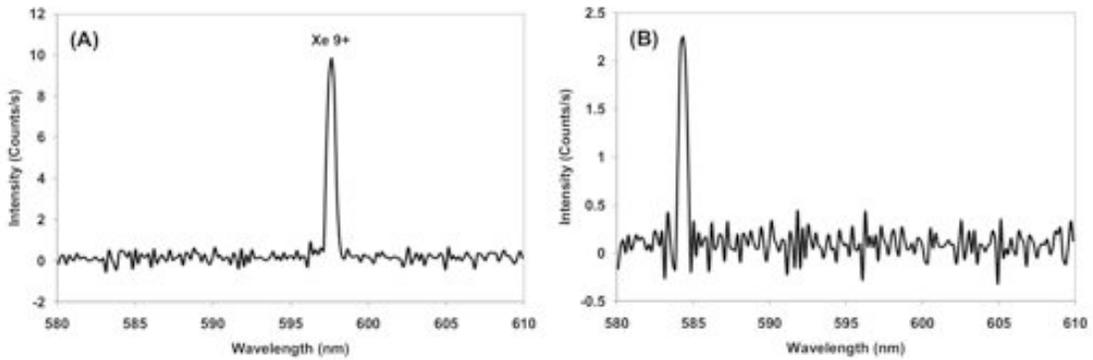

**FIGURE 3.** Portion of visible Xe spectrum, taken at beam energies of (A) 500 eV and (B) 100 eV.

# X-RAY MICROCALORIMETRY

With the award of the Nobel Prize in 2002 for the birth of x-ray astronomy, and the launch of the spectacular orbiting x-ray observatories Chandra and XMM Newton in 1999, it is an exciting time to be involved in x-ray astronomy. The present era has been likened to the early days of optical spectroscopy, when spectral lines visible to the eye were just being highly resolved for the first time. Future generations of x-ray observatories are being planned that will use new types of high-resolution spectrometers that are based on photon calorimetry rather than diffraction from a grating or crystal. Shortly before the launch of Chandra and XMM Newton, the Harvard-Smithsonian Astrophysical Observatory (Harvard-SAO) began a collaboration with us to test the development of x-ray microcalorimeters, and to deploy them for laboratory astrophysics using the NIST EBIT to simulate the hot plasmas that were soon to be studied by the orbiting observatories. The particular microcalorimeters and associated electronics deployed at NIST were built entirely at Harvard-SAO and Lawrence Berkeley National Lab in recent years (except for commercial dewars) and have not been deployed previously at other EBIT facilities.

Since the publication of our initial results on the astrophysically relevant Fe XVII spectrum [14], the Livermore EBIT group has also published some FeXVII results using a NASA-GSFC microcalorimeter [15]. The experimental results have stimulated significant theoretical activity, including that of Chen and Pradhan [16] who have carried out what is described as possibly the largest electron-ion scattering calculation to date (a relativistic close-coupling calculation with Hamiltonian matrix dimension up to 10,286). Prior to this calculation, the astrophysical observations were not comfortably reconcilable with either theory or our measurements (although some of the astrophysical line ratios did agree with the later results from the Livermore EBIT group [15]). The new calculations, however, show strong resonances along the energy axis and suggest that our measurements are actually the ones in better agreement with both theory and astrophysical observations, once the energy/temperature dependence is taken into account (figure 2 of reference [16]). Independent of the astrophysical observations, we can say that while some of the line ratios measured in the LLNL experiment are in relatively good agreement with our results, other ratios disagree at the 2-sigma to 3-sigma level. We are presently starting a new set of measurements to investigate the significance of this discrepancy further, as well as to study other spectra of interest.

Recently, both the LLNL and the NIST EBIT groups have deployed new (2nd generation) microcalorimeters in their labs. At the present conference, the LLNL group emphasized the performance improvements in their new microcalorimeter, as compared to our old instrument, so we point out here the corresponding improvements in our instrument as well. During a recent set of first measurements on the NIST EBIT, our new microcalorimeter obtained a resolution improved slightly from 6 eV [17] to 5.5 eV (3.0 eV at 6 keV photon energy has been obtained under more optimal conditions at SAO). The microcalorimeter at LLNL has improved from 11.5 eV to 6.5 eV (at a photon energy of 6.7 eV, a temperature of 60 mk, and for 14 summed pixels [18]). The total number of active pixels is 32 at LLNL [18] and 4 at NIST. Some of the early measurements at NIST used an x-ray lens that boosted the light intensity per pixel by up to a factor of 100 [19], rather than using more pixels to collect the naturally distributed light. The maximum count rate/pixel is limited by the pulse decay time, which is 1 ms in both instruments used at NIST, and has improved to 3 ms in the new instrument used at LLNL [20]. The pixel size is 0.624x0.624 mm$^2$ at LLNL [20] and 0.4x0.4 mm$^2$ at NIST. Our "hold time" (amount of time that it can be run continuously, before having to shut down and refresh the adiabatic demagnetization refrigerator), has improved from 10 hours to >24 hours, while the LLNL hold time has improved from 5 hours to >12 hours [18].

Because of the apparent discrepancy between the NIST and the LLNL Fe XVII measurements, a number of remarks have been published [15] by the LLNL group that require clarification. The most substantive ones are discussed below (this discussion is by the first author of the present paper, and should not necessarily be attributed to all of the other coauthors, only some of whom participated in the relevant experiments).

Contrary to the claim that the NIST EBIT suffers "performance limitations" and will not operate below 700 eV [15], figure 3 above demonstrates that it indeed operates quite well at energies as low at 100 eV. In fact, the NIST EBIT has operated significantly below 100 eV since the first time it was attempted, in January 2000. Like all EBITs (including the LLNL EBIT [18]), however, ours does show an increase in snout/anode current when operated at low beam energies [21]. This is typically compensated for by lowering the beam current.

It has been suggested [15] that the NIST EBIT operates at higher beam temperatures than the LLNL EBIT. There are two possible meanings of "beam temperature": (1) overall beam temperature, or full energy spread, and (b) transverse beam temperature. Regarding the first, the paper cited by LLNL in [15] concerning this issue determined their beam temperature to be 50 eV [22], *larger* than the 45 eV value determined previously for the NIST EBIT [23]. The actual value for the NIST EBIT during the Fe XVII experiment is likely to have been even less since we operated at reduced beam currents (one component of the temperature is proportional to the space charge, which scales linearly with beam current).

Regarding the second type of beam temperature, the relevant paper [24] referenced by the LLNL group in [15] quotes us [21] as saying that 700 eV is our estimate for the transverse beam temperature of the NIST EBIT, whereas in fact we only proposed that this was an upper bound. The LLNL group [24] specifies their own transverse beam temperature to be 190(30) eV (revised upwards from their earlier estimate of 110 eV [25]), reported to be in agreement with a theoretical prediction of 194 eV. Their theoretical prediction, however, is based on an imprecise value for one of the critical constants in the theory (the beam radius at the cathode). When the actual value [21, 26] for this constant is used, the theoretical prediction increases to 440 eV. Furthermore, the LLNL group obtained their experimental value of 190 eV from a set of determinations that ranged from –35 eV to +1920 eV (nearly 3 times our upper bound). They discarded half of these results, "even if the residuals intimate an excellent fit" [24], and averaged the remaining 6 points, even though these remaining points still showed a wide scatter (spanning 6-sigma of the final quoted uncertainty). It is clear from the details of their analysis that the spread in the 6 determinations is due primarily to a non-random systematic error arising from an improper form of the fitting function. It is therefore inappropriate that they quoted an uncertainty reduced below the standard deviation by the square root of the number of determinations [27]. If they had not discarded half of their determinations, they would have arrived at a mean value of 566 eV, in good agreement with the corrected theoretical prediction and much closer to our upper bound.

It has been claimed that the NIST EBIT does not operate under collisional equilibrium (citing evidence that Fe XVII lines do not appear in the LLNL measurements beyond 1.5 keV, whereas they do appear in the NIST measurements at these energies) [15]. The rapid approach to collisional equilibrium (<15 ms) has been demonstrated in figure 3 of our earlier work [28], for a range of charge states of Kr, from q=1 to q=22. In our Fe XVII experiment, the trapping time was 1.0 s [29], significantly longer than the time needed to reach collisional equilibrium in that system. The fact that Fe XVII lines do not appear in the LLNL measurements for beam energies beyond 1.5 keV can be easily explained by the lack of ion cooling gas in their experiment. Using an EBIT simulation code developed at LLNL [30], we have modeled the cases of no cooling gas, $P<5 \times 10^{-11}$ hPa (torr), and moderate amounts of cooling gas, $P=5 \times 10^{-9}$ hPa (torr), and found that in the former case the concentration of Fe XVII is reduced to zero above 1.5 keV, while in the latter case it is only rapidly reduced to a local minimum (still far above zero) around 4 keV, and then grows modestly up to (and beyond) 10 keV. We conclude that the difference in observed behavior between the LLNL and NIST EBIT is the normal result of charge exchange from the cooling gas, and does not imply a lack of collisional equilibrium in either one of the devices.

It was claimed in [15] that an earlier LLNL paper [31] proved that a substantial fraction of Fe XVI contaminated the NIST 3C/3D line ratio. In fact, the LLNL paper [31] showed only that Fe XVI contaminated some of their own work, not the NIST work. The only way the LLNL group was able to obtain significant contamination in their spectra was to continuously inject neutral Fe in a gaseous form. When they used pulsed MEVVA ion injection (as we did at NIST) the contamination was negligible. As pointed out by Chen and Pradhan [16], both the published LLNL and the NIST 3C/3D ratios are in good agreement with each other (and with theory) once the energy dependence is taken into account (at least at the energies calculated with resonances).

It was correctly pointed out [15] that the NIST Fe XVII data was not time-resolved. Indeed, we summed and reported every photon emitted during the normal course of the experiment. Although not described explicitly in their paper [15], the LLNL group discarded photons that were detected in the period just after iron injection, and analyzed only those that were detected after a certain delay time [18]. Although the difference is significant in principle, in practice it will not affect the final results much if the trapping time is long compared to the time to reach collisional equilibrium.

It was pointed out that our "calorimeter response was not calibrated or monitored in-situ" [15]. The response of our calorimeter was determined initially (including window transmission) and observed to be stable during the experiment. Because of the high overall stability, the much smaller differential changes in the response curve between nearby wavelengths should be entirely negligible. As calculated in our paper [14], it would take a huge window contamination layer, equal to twice our window thickness itself in the case calculated, to shift our most sensitive line ratios outside of our quoted error bars. We believe this is very unlikely.

Finally, we note a significant difference in the adjustment (or lack thereof) of the line ratios that were actually observed in the NIST and LLNL EBITs—a difference previously unappreciated by the LLNL group that appears to have led them to an incorrect comparison of one set of results to another in their figure 2 [15] and in their invited talk at the present conference [18]. As described in our papers, we report line ratios that are "as observed" (i.e. from a real EBIT under specified conditions)—there are no adjustments based on theory. The LLNL group, in contrast, adjusts their observed values (using theory) to obtain results that *would have been observed* if their EBIT were more similar to some astrophysical sources (for example, not having an electron beam that is unidirectional and oriented 90-degrees to the observation vector). In our view, such adjustments should be a part of the theory that explains the experimental data, not the other way around (compare our figure 3 [14], which shows the effect of the anisotropic emission in the theory, for a range of allowed possibilities, rather than as a fixed offset in the experiment). Once the two sets of data are plotted on equal footing, the apparent discrepancy is significantly reduced.

## CHARGE EXCHANGE STUDIES

With the surprising discovery of the emission of x-rays (typically corresponding to very high temperatures) from comets ("dirty snowballs"), and the subsequent explanation of this effect in terms of charge exchange with the solar wind, studies related to the capture of electrons from gasses by highly charged ions have become popular (for a review, see [32]). The LLNL EBIT was deployed to address the cometary x-ray problem using a technique in which the electron beam is switched off and low-level x-ray emission is observed to persist from the magnetically trapped ions due to charge exchange with the background gas inside the EBIT [33]. Motivated not only by the cometary x-ray problem, but also by the need for a better understanding of charge exchange processes in research involving controlled thermonuclear fusion and EUV lithography, we have been engaged in charge exchange studies using a complementary technique, since the year 2000. In our experiments [34-35], the range of charge states present in the EBIT are extracted and passed through a q/m selecting magnet, and then directed through an aperture into a differentially pumped chamber which contains a gas jet target. We observe the x rays emitted from the interaction of the ion beam with the jet, using a high purity Ge detector. This method is also complementary to the LLNL method in that ours involves collision velocities that typically are much higher (well-matched to those in the solar wind), although we can also adjust our beams to span the range of velocities from above 800 km/s to below 30 km/s. The velocity dependence of the measured and calculated "x-ray hardness factor" has been presented in figure 33 of a recent review article by Beiersdorfer [36]. Our results for Ne projectiles that are fully stripped before electron capture agree well with the experiment of Greenwood, et. al [37] shown in that plot [36].

In order to learn more about the details of the capture process, one can look at the full spectral signature of the x-ray emission. In figure 4, we show this for the case of Ar projectiles, now adding some complexity to the problem by leaving one electron on the projectile (before electron capture and emission in the He-like state).

We have been working with several sets of theorists to model this spectrum and gain insight into the quantum state-specific $(n, l)$ initial capture distribution. The result of several single-electron capture distributions calculated for our argon projectiles with the Classical Trajectory Monte Carlo (CTMC) method, following [38], and a version of the Landau-Zener (MCLZ) method, following [39], are shown in figure 5 (both include capture only from the Ar 3p shell). These distributions are used as input to a separate calculation, following [40], of the photon emission that occurs as the excited ion cascades to the ground state. The final results is shown in the curves plotted in figure 4. The discrepancy with the LZ model at the highest x-ray energies might be due to the neglect of multiple electron capture, anisotropic emission (polarization effects), or variation in the triplet-singlet population ratio (among other possibilities). We are currently engaged in a dynamic interplay between experiment and theory in attempt to reduce the disagreements and improve the understanding of the underlying physical processes in these and the other studies described in this paper.

A variety of other experiments are under way at the NIST EBIT facility, including continued studies in the area of ion-surface interactions [41] and precision measurements of the effect of the quantum vacuum on atomic structure [42].

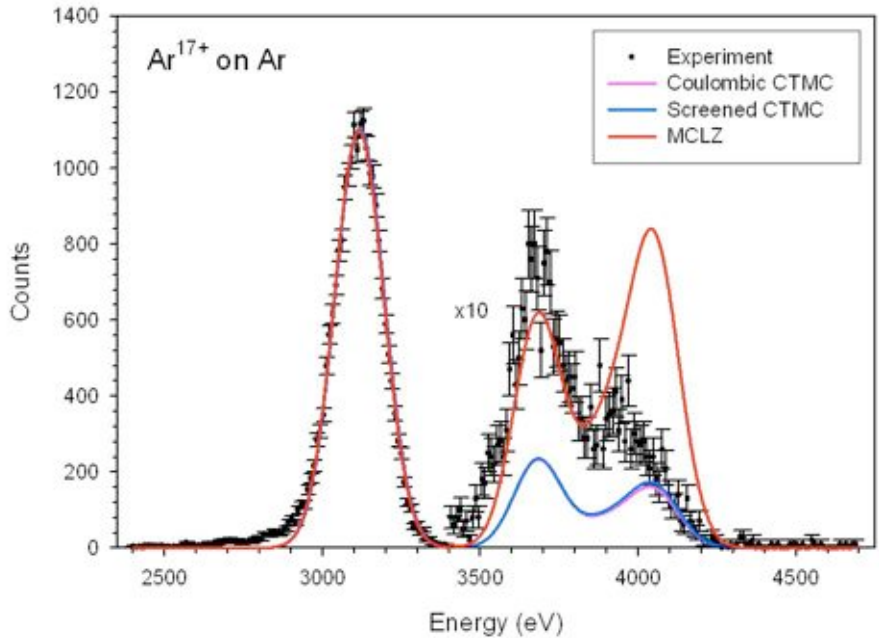

**FIGURE 4.** X-rays observed at 90° after 800 km/s Ar$^{17+}$ ions capture electrons from Ar gas.

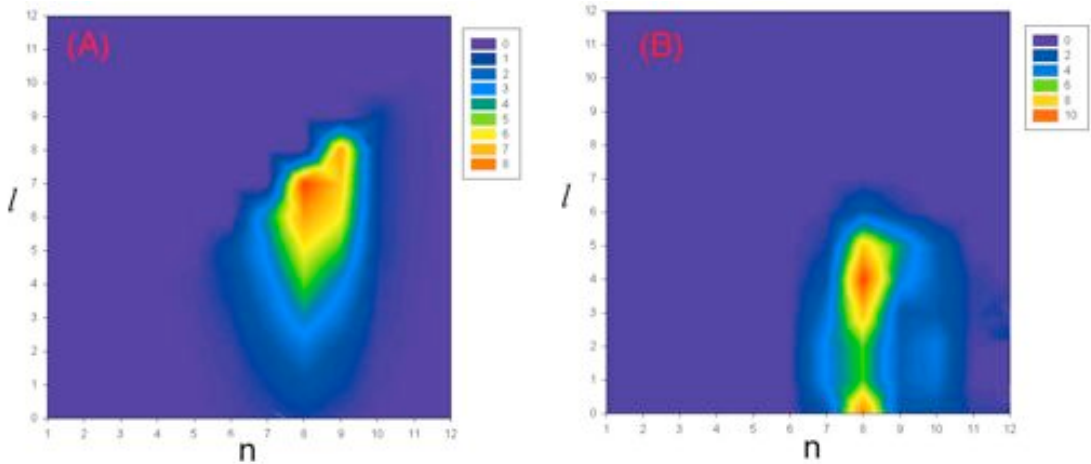

**FIGURE 5.** Initial capture probability distributions as a function of principal quantum number (n) and orbital angular momentum (l), calculated with (A) screened CTMC [38] and (B) MCLZ [39] methods.

## ACKNOWLEDGMENTS

We thank the following collaborators/coauthors who have helped in various aspects of the three types of experiments described in this paper: A. Aguilar, D. J. Alderson, G. Austin, A. Bhatia, M. Barbera, J. Beeman, J.-P. Briand, N. Brickhouse, J. Burnett, G. Doschek, E. Haller, Q. Kessel, I. Kink, D. Landis, N. Madden, K. Makonyi, S. Murray, G. O'Sullivan, J. E. Pollack, E. Parra, Pomeroy, J. R. Roberts, W. Smith, E.

Sokell, and C. Verzani. We thank DOC, NASA, DOE, and ISMT for funding this work at the NIST EBIT facility. In addition, V.K. was supported by the NASA grant NNG04GD57G and A.D. by the NASA grant NAG5-13331; E.T. was partially supported by the Hungarian Science Fund (OTKA) T046454 and T042729.